\documentclass[11pt]{article}

\usepackage[margin=1in]{geometry}
\usepackage{booktabs}
\usepackage{float}
\usepackage[hidelinks]{hyperref}
\usepackage{natbib}
\usepackage{graphicx}
\usepackage{amsmath}
\usepackage{pgfplots}
\usepackage{tabularx}
\pgfplotsset{compat=1.18}
\definecolor{accentblue}{HTML}{2A5599}

\title{Uneven Self-Reported AI Use: An Embedded Case Study Across Eleven Teams in a Software Organization}
\author{Daniele Verì}
\date{November 10, 2026\\[4pt]\small\href{https://arxiv.org/licenses/nonexclusive-distrib/1.0/license.html}{arXiv.org non-exclusive license to distribute (v1.0)}}
\hypersetup{
  pdftitle={Uneven Self-Reported AI Use: An Embedded Case Study Across Eleven Teams in a Software Organization},
  pdfauthor={Daniele Verì}
}

\begin{document}
\maketitle

\begin{abstract}
We report a descriptive embedded single-case study \citep{yin2018} of self-reported AI use in a software organization. The analysis retains 98 scored interview records: 65 collected with ten standardized team-level instruments and 33 with one distinct team-level instrument. On the standardized-form base, 36/65 respondents (55.4\%) reported a numeric frequency score of 4 or 5; across the ten teams using standardized forms, scores range from 1/5 (20\%) to 5/7 (71.4\%). In a record-level cross-tab, 31/36 (86.1\%) respondents scoring 4--5 and 28/29 (96.6\%) scoring 0--3 reported no policy awareness or unclear policy details. The distinct-form group produced 31/33 (93.9\%) at the same numeric cutpoint; the pooled 67/98 (68.4\%) numeric summary is secondary because its verbal anchors differ. These counts describe the interviewed sample, rather than organizational prevalence, policy compliance, or the value of AI use. Interviewer-entered notes illustrate reported constraints and practices without establishing their prevalence or explaining the frequency differences. The primary contribution is a descriptive organizational case, supported by transparent documentation of the comparisons the available materials permit and their limits.
\end{abstract}

\section{Introduction}

Teams using large language model (LLM) based tools may face different tasks, data-sensitivity constraints, and histories of tool access. An aggregate frequency percentage describes the interviewed sample; team counts provide additional information about variation within that sample. Neither alone measures the maturity or value of AI use.

The organization commissioned an internal study of its AI use. Instruments were adapted during collection, sometimes through unit-specific extensions; one unit used an earlier, structurally distinct instrument. The resulting eleven forms differ in content and response structure. These differences limit the comparisons available in an external account of the study.

Following \citet{yin2018}, we use the company as a single case and its eleven teams as embedded units, retaining the standalone record in the overall sample. The empirical descriptive case is the primary contribution; the item-compatibility review is supporting documentation. The exploratory questions guiding this account are: how does reported frequency vary across teams in the interviewed sample; what practices and constraints do respondents describe; and which comparisons does the available instrument documentation permit? These questions describe the present analytical scope and are not presented as preregistered questions.

\section{Related Work}

Case study methodology for organizational research, including the embedded single-case design used here (one organization, multiple sub-units of analysis), follows \citet{yin2018}. The request for a recent concrete task was inspired by the Critical Incident Technique \citep{flanagan1954}. The materials do not document systematic application of its criteria for situations, behavior, and consequences, so we do not claim a full application of that technique or direct observation of behavior.

Two empirical studies provide useful comparison points. In a staggered field deployment involving 5,172 customer-support agents, generative-AI assistance produced heterogeneous productivity effects across experience and skill levels \citep{brynjolfsson2023}. An interview study of engineers and collaborators in integrated-circuit design and other high-precision work instead emphasizes accuracy concerns and the difficulty of controlling the context in which AI tools operate \citep{moss2025}. These studies do not benchmark the present case: the first measures a deployed intervention and the second studies a high-precision engineering setting. The present contribution is narrower—an internal, across-team description of reported frequency, policy awareness, and reported work practices in one organization, with explicit separation of instrument conditions.

\section{Methods}

\subsection{Setting}

The case is a privately held software organization. The organization commissioned this study internally; one staff member conducted every interview. The organization's name and products are omitted, and no individual is named.

The scored interviews and interviewer-administered scoping interviews were live video calls between one interviewer and one respondent, lasting approximately 10--30 minutes, collected in July--September 2026. Other structured responses were entered during the call, while open-text answers were sometimes paraphrased or condensed. No separate audio recording or post-call transcription was used. The executive-sponsor questionnaire was self-administered and is excluded from this description of live calls. Completed interview summaries were not returned to respondents or managers for review or correction.

\subsection{Instrument design}

Eleven team-level Tally instruments were structured, interviewer-administered questionnaires. Ten share a core of roughly 14--16 items covering tool access and use, self-reported frequency on a 0--5 scale, a recent concrete task, perceived effects, tasks where AI performs poorly, daily tools, sensitive data, policy awareness, training needs, and additional notes. The current question text and form metadata were reviewed on 24 September 2026. At that time, the listed forms' last-update metadata all preceded the review, and no later form update was recorded. Appendix~\ref{app:items} presents the two quantitative items used in the analyses.

\pagebreak

One unit used an earlier questionnaire that asks about configuration practices, fallback tools, and execution capabilities, with a distinct item set.

Beyond this shared core, four instruments carry extensions that are not present on the other forms. These cover agentic (autonomous-execution) use and rate limits; platform preference, contextual memory setup, and public-versus-internal task framing; and tool-selection criteria. The earlier questionnaire also differs in its item set.

Scoping respondents were interviewed before the corresponding team interviews using a distinct Tally instrument administered by the same interviewer. The available scoping materials identify twelve records: eleven from team-level respondents and one cross-team respondent. We retain this separate base of twelve records, whose qualitative responses and multiple-selection priorities informed instrument design and the descriptive results. An additional questionnaire was self-administered in a different language from the team forms and is outside both bases. Person-level overlap between scoping respondents and the 98 scored respondents could not be established, so we do not claim a total number of unique people across the two bases.

\subsection{Item compatibility and coverage review}
\label{sec:item-audit}

Table~\ref{tab:item-audit} summarizes item compatibility and coverage based on the current Tally question text reviewed on 24 September 2026. This review concerns documented content, response structure, and analytical coverage; it does not establish psychometric equivalence or measurement invariance. Matching wording alone would not establish that respondents interpreted a measure identically.

The frequency item, ``How often do they use AI?'', is reported across all eleven instruments. The current versions of the ten standardized forms share the same recorded anchor labels at the left endpoint, midpoint, and right endpoint of the 0--5 scale: 0 ``Rarely,'' 3 ``Few times a week,'' and 5 ``Most working days.'' The current distinct-form instrument uses the same numeric positions but different labels: 0 ``Rarely,'' 3 ``Few times / Week,'' and 5 ``More times / Days.'' These differences in current recorded labels do not by themselves establish that participants interpreted the scales differently. We retain the 98-record counts as a secondary numeric summary while treating their cross-instrument interpretation as conditional; the labels do not establish measurement invariance.

The open-text task and difficulty items do not have fixed response options in this account. Their surrounding prompts, context, and administration are not fully documented, so the qualitative material is not treated as a set of identical coded response categories. As a concrete compatibility example, selecting an execution capability in the distinct form's multi-select list cannot be mapped to the four-option choice between unsupervised, approval-gated, no, and unsure in another form: the former does not establish a level of human control. The two counts therefore remain separate in Table~\ref{tab:agentic}.

\begin{table}[H]
\centering
\caption{Item-compatibility and coverage summary. Entries describe available documentation, not reconstructed exact item text.}
\label{tab:item-audit}
\footnotesize
\begin{tabularx}{0.98\textwidth}{@{}>{\raggedright\arraybackslash}p{2.1cm}>{\raggedright\arraybackslash}p{1.9cm}>{\raggedright\arraybackslash}X>{\raggedright\arraybackslash}X@{}}
\toprule
Item family & Coverage & Known anchors or response structure & Analytical decision \\
\midrule
Use frequency & Eleven instruments; 98 records & Participant-stated 0--5 score; current anchors at 0, 3, and 5 verified across ten standardized forms; one form uses distinct wording & Primary comparison within the 65 standardized records; 67/98 retained as a secondary arithmetic summary \\
Recent task; AI performs poorly & Reported on all eleven instruments & Open text with similar stems; surrounding prompts and their influence are not fully documented & Illustrative paraphrases and descriptive categories; no cross-team category prevalence \\
Policy awareness & Standardized forms; n=65 & ``Yes, and follows it,'' ``Aware but unclear on details,'' ``Not aware of any policy'' & Report 6/65, 18/65, 41/65 as self-reports; no inference about policy existence or compliance \\
Execution and human control & Two forms with different response structures & Four-option single choice versus multi-select capability list & Report items separately; no organization-wide agentic rate \\
Scoping priorities & Separate scoping base; 12 records & Multiple-selection items; selections may overlap & Retain as separate contextual results; no extension to the scored sample \\
\bottomrule
\end{tabularx}
\end{table}

\subsection{Sample and scoring}
\label{sec:sample}

The scored base comprises 98 interview records collected using eleven team-level instruments. Ten standardized-form teams contribute 64 records, and one team using a distinct form contributes 33; one additional standardized-form record is retained as a standalone case and is not assigned to a team. The eleven team-level groups total 97 records, including 66 with numeric scores of 4 or 5. Including the standalone record gives 98 records and 67 scores at that cutpoint.

The primary cross-team comparison uses the 65 records collected with the ten standardized instruments, including the standalone record in the form-level base. The ten teams using standardized forms contribute 64 records to the chart; the standalone record is excluded from team-level plotting. The distinct-form team is retained as a separate instrument group rather than treated as an equivalent measurement condition.

This is a non-probability sample of team members available and willing to be interviewed during July--September 2026. The team-by-team chronology and any contemporaneous changes in access, policy, or training are not documented, and no validated response-rate denominator or proximity to a census can be established. Percentages describe the interviewed sample. All team percentages should be read with their counts, including groups of five or seven: one response changes a percentage by 20 points for a group of five. The standalone record is not a team.

We refer to scores of 4 or 5 as the high-frequency group for this analysis, subject to the cross-instrument anchor differences in Section~\ref{sec:item-audit}. A score below this threshold does not establish non-use: lower-frequency use can still involve substantial integration or valuable work. Frequency alone measures neither autonomy nor value. The full numeric score distribution and sensitivity to alternative numeric cutpoints are reported in Section~\ref{sec:sensitivity}; these cutpoints do not resolve the non-equivalence between the verbal anchors of the distinct form and those of the standardized forms.

For the frequency--policy cross-tab, we used the 65 standardized-form records with both items answered, grouped scores as 0--3 or 4--5, and combined ``Aware but unclear on details'' with ``Not aware of any policy'' as ``unaware or unclear.'' ``Yes, and follows it'' remained a separate response.

\subsection{Qualitative analysis}
\label{sec:qualitative}

The available qualitative material consists of interviewer-entered live notes, including paraphrase and synthesis of spoken answers. We treat each recorded response or described task as a descriptive unit and organize it under content categories such as data handling, policy uncertainty, training requests, configuration practices, and automation. These categories summarize the content available in the notes; they are not validated constructs or evidence of thematic saturation. Examples are selected to illustrate a reported practice, concern, or request, rather than to represent its frequency within a team or across the company.

The interviewer also prepared the descriptive synthesis. There was no independent second coder or respondent validation. The available materials do not document a formal codebook, successive coding rounds, systematic searches for disconfirming cases, or a complete note-to-finding audit trail. We conducted no additional qualitative coding for this analysis. Examples below are paraphrases or analytical summaries, not verified verbatim quotations; their selection may reflect the interviewer's perspective. Differences in prompts and coverage preclude interpreting an unmentioned concern as absent from a team.

\subsection{Anonymization and reporting}

Teams are designated Team 1 through Team 11 to protect confidentiality. The company and its products are not named. Qualitative accounts are paraphrased and generalized, and combinations of role and contextual details that could identify a participant are omitted.

\section{Results}

\subsection{Primary frequency comparison uses standardized instruments}

Within the 65 standardized-form records, 36 (55.4\%) have a participant-stated numeric score of 4 or 5. Across the ten teams using standardized forms, charted percentages range from 1/5 (20\%) to 5/7 (71.4\%); the standalone record is not assigned to a team. The distinct-form team has 31/33 (93.9\%) at the same numeric cutpoint. This difference is reported as a contrast between instrument groups, not as a fully equivalent behavioral measure.

As a secondary arithmetic summary, 67/98 records (68.4\%) have numeric scores of 4 or 5: 31/33 in the distinct-form group and 36/65 in the standardized-form base. Because the verbal anchors differ, this pooled figure is not a common behavioral measure across instruments.

\subsection{Full score distribution and numeric threshold sensitivity}
\label{sec:sensitivity}

Table~\ref{tab:score-distribution} reports the complete numeric distribution of frequency scores across the 98 interview records. Scores cluster near the top in the distinct-form group; the standardized-form group is more dispersed. These counts describe recorded scores; they do not make the distinct form's verbal anchors equivalent to those of the ten standardized forms. The pooled table is therefore a secondary numeric sensitivity display, not the primary cross-team comparison.

\begin{table}[H]
\centering
\caption{Distribution of self-reported frequency scores on the recorded 0--5 scale.}
\label{tab:score-distribution}
\small
\begin{tabular*}{\textwidth}{@{\extracolsep{\fill}}lrrrrrrr@{}}
\toprule
Scope & $N$ & 0 & 1 & 2 & 3 & 4 & 5 \\
\midrule
Distinct-form group & 33 & 1 & 0 & 0 & 1 & 1 & 30 \\
Standardized-form group & 65 & 11 & 3 & 6 & 9 & 6 & 30 \\
Overall & 98 & 12 & 3 & 6 & 10 & 7 & 60 \\
\bottomrule
\end{tabular*}
\end{table}

As a numeric sensitivity check, a cutpoint of 3 or higher gives 77/98 (78.6\%), 32/33 (97.0\%) in the distinct-form group, and 45/65 (69.2\%) in the standardized-form group. The 4-or-higher cutpoint gives 67/98 (68.4\%), 31/33 (93.9\%), and 36/65 (55.4\%); a cutpoint of 5 gives 60/98 (61.2\%), 30/33 (90.9\%), and 30/65 (46.2\%). These are numeric cutpoint sensitivities, not evidence that the verbal categories have equivalent meanings across forms.

\subsection{Team-level frequency scores within the standardized-form group}

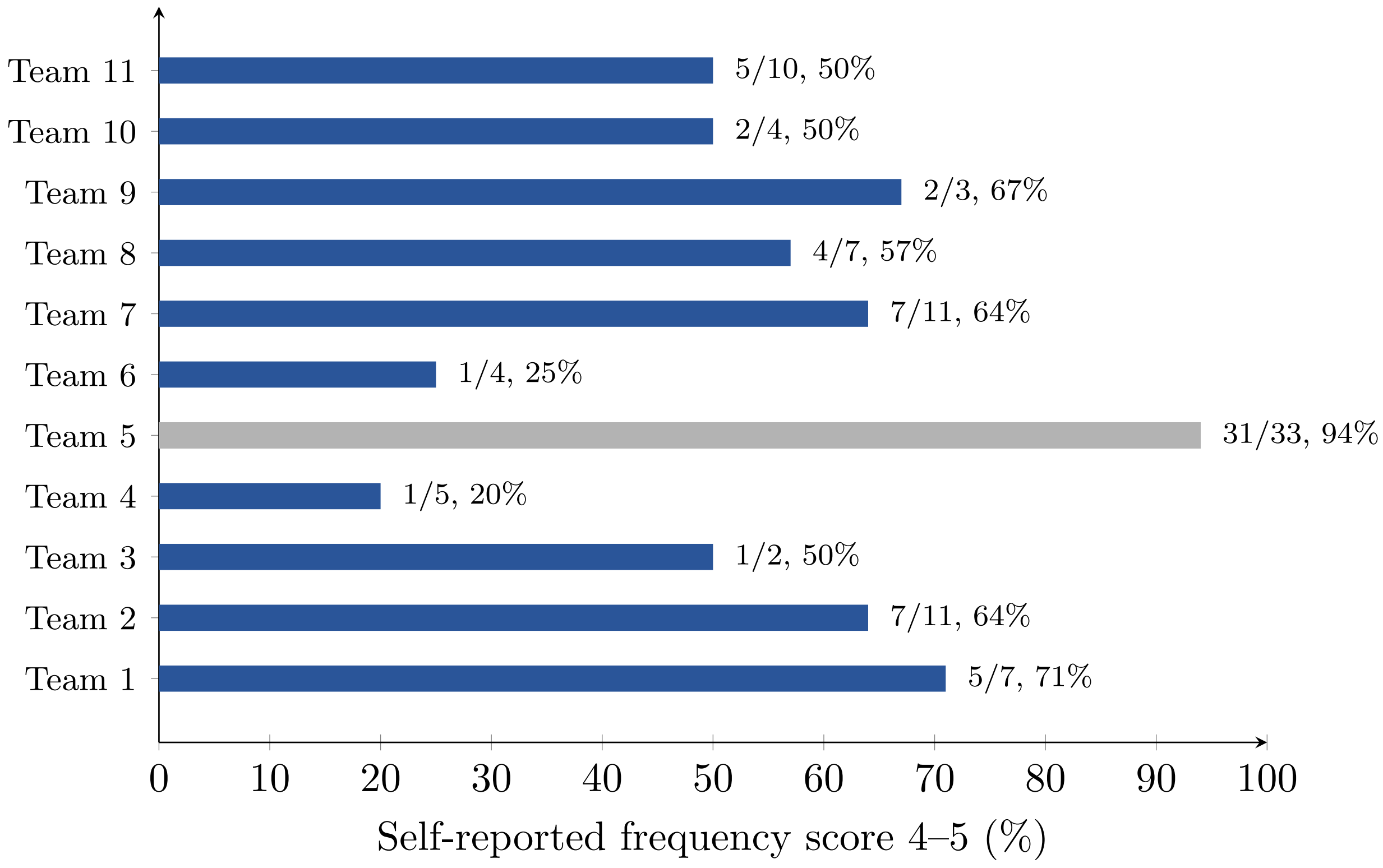
\begin{figure}[H]
\centering
\resizebox{\textwidth}{!}{%
\begin{tikzpicture}
\begin{axis}[
    xbar,
    xmin=0,
    xmax=100,
    width=12cm,
    height=8.5cm,
    bar width=7pt,
    bar shift=0pt,
    xlabel={Self-reported frequency score 4--5 (\%)},
    symbolic y coords={Team 1,Team 2,Team 3,Team 4,Team 5,Team 6,Team 7,Team 8,Team 9,Team 10,Team 11},
    ytick={Team 1,Team 2,Team 3,Team 4,Team 5,Team 6,Team 7,Team 8,Team 9,Team 10,Team 11},
    y tick label style={font=\footnotesize},
    point meta=explicit symbolic,
    nodes near coords={\pgfplotspointmeta},
    nodes near coords style={font=\scriptsize, anchor=west, xshift=2pt},
    axis y line=left,
    axis x line=bottom,
    enlarge y limits={abs=0.6cm},
]
\addplot[fill=accentblue, draw=none] coordinates {
(71,Team 1) [5/7, 71\%]
(64,Team 2) [7/11, 64\%]
(50,Team 3) [1/2, 50\%]
(20,Team 4) [1/5, 20\%]
(25,Team 6) [1/4, 25\%]
(64,Team 7) [7/11, 64\%]
(57,Team 8) [4/7, 57\%]
(67,Team 9) [2/3, 67\%]
(50,Team 10) [2/4, 50\%]
(50,Team 11) [5/10, 50\%]
};
\addplot[fill=gray!60, draw=none] coordinates {
(94,Team 5) [31/33, 94\%]
};
\end{axis}
\end{tikzpicture}%
}
\caption{Respondents reporting numeric frequency scores 4--5, shown as count/base and rounded percentage for each team. Gray marks the team surveyed with a questionnaire whose verbal anchors differ from the ten standardized forms. The standalone record is excluded from the chart. Read all percentages with their denominators: one response changes a team percentage by between approximately 3 and 50 points. The scale-compatibility limits in Section~\ref{sec:item-audit} apply.}
\label{fig:adoption-by-team}
\end{figure}

Within the ten teams represented by standardized forms, threshold percentages range from 1/5 (20\%) to 5/7 (71.4\%). The distinct-form team is shown separately at 31/33 (94\%, rounded). Figure~\ref{fig:adoption-by-team} shows all eleven team counts and rounded percentages, with the distinct-form team shaded separately. The standalone record contributes to the standardized-form base but is excluded from the team chart. Small counts throughout the chart limit interpretation of its ordering (Section~\ref{sec:sample}).

\subsection{Reported constraints across work contexts}

The interview notes contain concerns and requests in several work contexts. The following paraphrases illustrate reported constraints; differing prompts and coverage prevent conclusions about their distribution or their role in explaining frequency differences:

\begin{itemize}
\item Participants described concerns about trusting AI outputs and handling information with external providers.
\item The notes include uncertainty about whether policy permitted using sensitive work information with AI tools. This does not establish whether a relevant policy existed.
\item Participants requested tools suitable for sensitive information and additional training. These requests do not establish a gap in existing training, which was not assessed.
\end{itemize}

These examples do not support a pooled statistic on reasons for non-use or below-threshold use, nor do they demonstrate that any concern is unique to a team.

\subsection{Policy awareness}

The current Tally submissions report responses to the policy-awareness item for the 65 scored respondents in the standardized-form group: 41/65 selected ``Not aware of any policy,'' 18/65 selected ``Aware but unclear on details,'' and 6/65 selected ``Yes, and follows it.'' Thus 59/65 (91\%) reported either no awareness or unclear details. This measures reported awareness, not whether a policy existed or whether respondents complied with it; the distinct-form instrument did not ask this item.

Table~\ref{tab:frequency-policy} cross-tabulates frequency score bands and reported policy awareness for the 65 complete standardized-form records. Both items were answered in all 65 submissions. The table shows that 31 of the 36 respondents with a numeric frequency score of 4 or 5 were unaware or unclear (86.1\%), while 5 selected ``Yes, and follows it.'' Among the 29 respondents scoring 0--3, 28 were unaware or unclear (96.6\%) and 1 selected ``Yes, and follows it.'' Thus the reported uncertainty is substantial in both score bands. This is a descriptive co-occurrence, not evidence of policy violation, an effect of policy awareness, or a causal relationship.

\begin{table}[htbp]
\centering
\caption{Numeric frequency cutpoint by reported policy awareness in the 65 standardized-form submissions.}
\label{tab:frequency-policy}
\small
\begin{tabular*}{\textwidth}{@{\extracolsep{\fill}}lrrr@{}}
\toprule
Policy-awareness response & Score 0--3 & Score 4--5 & Total \\
\midrule
Unaware or unclear & 28 & 31 & 59 \\
Yes, and follows it & 1 & 5 & 6 \\
Total & 29 & 36 & 65 \\
\bottomrule
\end{tabular*}
\end{table}

\clearpage

\subsection{Reported practices, execution capabilities, and human control}

The notes describe shared configuration, a recurring automated workflow, and cross-model checking. One note describes a shared configuration practice but does not identify a specific output or its quality. Another describes a recurring automated workflow but does not consistently specify the delegated action, approval point, or human intervention. A third describes using one model to check another's output but does not establish an improvement in correctness. These are reported practices, not independently observed behavior, and they illustrate why frequency, integration, and autonomy need to be considered separately.

Responses to the distinct-form questionnaire included the multi-select item ``Can the agent execute actions?''; 32/33 respondents selected at least one execution capability, described as editing files, running commands, or opening changes for review. The item does not distinguish a capability available in a tool from one used by the respondent, or establish how much supervision accompanied it. A separate four-option question about execution and human control was used in another form. Table~\ref{tab:agentic} retains the two results separately; neither supports an organization-wide agentic-use rate. The 32/33 capability responses and 31/33 frequency-threshold responses within the distinct-form group measure different aspects and need not coincide.

\begin{table}[htbp]
\centering
\caption{Execution-related self-reports from differently structured items, reported separately (Section~\ref{sec:item-audit}). Category labels summarize the available account rather than reproduce exact questions.}
\label{tab:agentic}
\begin{tabularx}{\textwidth}{@{}p{4cm}X@{}}
\toprule
Response structure & Result \\
\midrule
Single-choice item & 1/10 unsupervised, 3/10 approval-gated, 2/10 no, 4/10 unsure \\
Capability multi-select & 32/33 report at least one execution capability \\
\bottomrule
\end{tabularx}
\end{table}

Other notes describe a scheduled workflow and internally developed tools. These descriptions do not establish autonomous AI execution or specify consistently which actions AI performs, where approval is required, or how humans monitor and intervene; their autonomy, reliability, and prevalence remain unverified.

\subsection{Scoping priorities and contextual correspondences}
\label{sec:scoping-priorities}

The separate scoping Tally submissions contain multiple-selection priorities: requests for training and questions about security/privacy were each selected by 6/12 respondents, questions about company rules by 5/12, and the absence of a shared team position by 4/12. Requests for company-provided tools were selected by 4/12 and requests for access to specific tools by 3/12. These counts describe twelve scoping responses, not the 98 scored staff interviews or twelve independent team assessments; selections overlap.

Comparing the available summaries also yields examples of related content: training and redaction requests alongside data-handling concerns; data-handling rules alongside deliberate limits on use; local-tool requests alongside trust concerns; and shared context and review requests alongside reported configuration and checking practices. These are descriptive correspondences, without a pooled agreement rate, confirmation that needs were unmet, or evidence that the proposed interventions would change frequency.

\section{Discussion}

The primary contribution is a descriptive case of participant-stated AI frequency across teams in one organization. Within the standardized-form base, 36/65 (55.4\%) respondents scored 4 or 5, and the ten teams using those forms range from 1/5 (20\%) to 5/7 (71.4\%). The same base shows a frequency--policy co-occurrence: 31/36 (86.1\%) respondents scoring 4--5 and 28/29 (96.6\%) scoring 0--3 reported no policy awareness or unclear details. Most respondents in each score band reported no awareness of a policy or uncertainty about its details. The distinct-form group yields 31/33 (93.9\%) at the same numeric cutpoint, while the pooled 67/98 (68.4\%) is a secondary arithmetic summary. These results make the composition of the interviewed sample visible, but establish neither organizational prevalence nor a typical team's experience.

The review of the forms and response structures documents coverage and response-structure differences; it is not a validation of the measures. The reported constraints and workflow descriptions add context, but cannot explain the quantitative differences. Task mix or familiarity with AI tools may be relevant interpretations of the variation, but they were not tested. Team affiliation, collection timing, instrument version, access, and task mix have not been separated analytically. Compared with field-deployment studies that measure productivity and interview studies focused on context control in high-precision work, this case contributes a narrower internal, cross-team description rather than an effect estimate or outcome evaluation \citep{brynjolfsson2023,moss2025}.

The scoping responses identify training, security/privacy, company rules, shared team position, and access to company-provided or specific tools as recurring priorities. These are participant-informed priorities for subsequent action, not evaluated interventions. Existing internal practices offer examples to examine and share, but this study does not establish their effectiveness or a causal effect on adoption.

\section{Limitations}

The scored frame comprises 98 records across eleven teams plus one standalone record. Ten teams used standardized forms and one used a distinct form. A separate scoping base contains twelve responses; its person-level overlap with the scored records remains unresolved. Scoping priorities are reported only for that separate base and are not extended to the scored sample.

All frequency results are self-reports without independent behavioral validation. The ten standardized forms have matching anchor definitions in the current form review, while the distinct-form instrument uses different verbal anchors; the full numeric score distribution and alternative numeric cutpoints are reported in Section~\ref{sec:sensitivity}, but they do not establish semantic equivalence across that boundary. The policy item also combines knowledge and declared adherence in ``Yes, and follows it'' and offers no separate response for knowing a policy but not following it; the table therefore does not measure knowledge independently of declared compliance. The willing-and-available sample has no validated response-rate denominator; small denominators limit interpretation of differences and rankings. Lower-frequency use cannot be classified as non-use from the threshold counts.

Collection timing and instrument version are intertwined: one unit used an earlier distinct instrument during the July--September 2026 collection window. We could not reconstruct the unit-by-unit chronology or establish whether access, policy, or training changed during data collection. These descriptive data cannot separate stable between-unit differences from collection conditions or change over time.

The qualitative results reflect one interviewer's live notes and synthesis, without recordings, independent second coding, respondent validation, or a complete analytical audit trail. The separate scoping base is not matched to the scored records and cannot support respondent-level comparisons with standardized criteria. Internal commissioning and the interviewer's staff role may have shaped responses, especially on sensitive topics; the available data cannot quantify this influence.

\section{Ethics, consent, and researcher role}
\label{sec:ethics}

The study was originally conducted as an internal organizational research initiative. Participants were informed about the internal research purposes, while external academic publication was not part of the original research plan. The decision to prepare an academic manuscript was made after data collection.

The organization reviewed and authorized the external publication of the anonymized research findings. The published manuscript contains aggregated quantitative results and paraphrased qualitative accounts. Direct personal identifiers, company and product names, and distinctive individual details were excluded to reduce identification risks.

The researcher conducted the interviews as an employee of the organization, as part of their professional responsibilities. This dual role may have influenced participants' responses, particularly on sensitive topics such as organizational policies and data-handling practices.

No individual-level interview records or identifiable participant responses are included in the publication. Small team sizes and combinations of contextual information may nevertheless present residual identification risks, which informed the selection, aggregation, and presentation of findings.

\section{Conclusion}

The primary result is a descriptive comparison within the 65 standardized-form records: 36 (55.4\%) have participant-stated numeric scores of 4 or 5, with the ten teams using standardized forms ranging from 1/5 (20\%) to 5/7 (71.4\%). The frequency--policy cross-tab shows that 31/36 (86.1\%) respondents scoring 4--5 and 28/29 (96.6\%) scoring 0--3 reported no policy awareness or unclear details; the uncertainty is substantial in both bands and does not measure policy existence or compliance. The distinct-form group yields 31/33 (93.9\%) at the same numeric cutpoint, while the secondary pooled arithmetic is 67/98 (68.4\%). These findings support a descriptive account of recorded frequency, policy-awareness responses, and illustrative work practices in one organization. The item-compatibility and coverage review supports interpretation of that case rather than establishing a validated measurement method; the results should not be read as a common behavioral scale across all instruments.

\clearpage
\appendix
\section{Quantitative item wording and response options}
\label{app:items}

\paragraph{Frequency item.} ``How often do they use AI?'' Linear scale from 0 to 5. The ten standardized forms use 0 ``Rarely,'' 3 ``Few times a week,'' and 5 ``Most working days.'' The distinct-form instrument uses the same numeric positions with 0 ``Rarely,'' 3 ``Few times / Week,'' and 5 ``More times / Days.''

\paragraph{Policy-awareness item.} ``Is the participant aware of any policy on AI tool use and data handling?'' Response options: ``Yes, and follows it''; ``Aware but unclear on details''; ``Not aware of any policy.''

\bibliographystyle{plainnat}

\end{document}